\documentstyle[11pt,Cargesepasp,twoside,epsf]{article}
\def\edcomment#1{\iffalse\marginpar{\raggedright\sl#1\/}\else\relax\fi}
\reversemarginpar

\begin{document}
\title{Star Formation from Turbulent Fragmentation }
 \author{Ralf Klessen}
\affil{Sterrewacht Leiden, Postbus 9513, 2300-RA Leiden, The Netherlands}

\begin{abstract}
  Star formation is intimately linked to the dynamical evolution of
  molecular clouds. Turbulent fragmentation determines where and when
  protostellar cores form, and how they contract and grow in mass via
  accretion from the surrounding cloud material. Efficiency, spatial
  distribution and timescale of star formation in turbulent clouds are
  estimated by comparing numerical models of self-gravitating
  isothermal gas where turbulence is assumed to have decayed or is
  driven at supersonic speed on various wavelengths.  Turbulence that
  is not continuously replenished or that is driven on large scales
  leads to a rapid formation of stars in a clustered mode, whereas
  interstellar turbulence that carries most energy on small scales
  results in isolated star formation with low efficiency.
  Protostellar accretion rates strongly vary in time and peak at a few
  $10^{-5}\,$M$_{\odot}\,$yr$^{-1}$. The clump mass spectrum for
  models of pure hydrodynamic turbulence is steeper than the observed
  one, but becomes comparable when including gravity. The mass
  spectrum of dense cores, on the other hand, is log-normal for
  decaying and large-wavelength turbulence, similar to the IMF, but is
  to flat in the case of small-scale turbulence.
\end{abstract}

\section{Introduction}
Understanding the processes that lead to the formation of stars is one
of the fundamental challenges in astronomy.  Stars are born in
turbulent interstellar clouds of molecular hydrogen. The location and
the mass growth of young stars are hereby intimately coupled to the
dynamical cloud environment. Stars form by gravitational collapse of
shock compressed density fluctuation generated from the supersonic
turbulence ubiquitously observed in molecular cloud (e.g.\ Padoan
1995, Klessen, Heitsch, \& Mac~Low 2000). Once a gas clump becomes
gravitationally unstable, collapse progresses in the center until a
protostellar object forms which continues to grow in mass via
accretion from the infalling envelope (see Wuchterl, this volume).
Also, stars hardly ever form in isolation. Instead they come in small
aggregates or larger clusters (Lada 1992, Mizuno et al.\ 1995, Testi,
Palla, \& Natta 1998), where the interaction of protostellar cores and
their competition for mass from their surrounding are important
processes shaping the distribution of the final star properties.

Star formation can thus be seen as a two-phase process: First, {\em
  turbulent fragmentation} leads to transient clumpy molecular cloud
structure, with some of the density fluctuation exceeding the critical
mass and density for gravitational contraction. This sets the stage
for the second phase of star formation, the {\em collapse of
  individual protostellar clumps} building up central protostars. In
the following, the words gas clump and shock-generated density
fluctuation are used synonymously. Cores are
defined as the high-density central regions of collapsing clumps where
protostars build up. I focus  on the first phase of star formation and
demonstrate how the turbulent interstellar velocity field 
influences the properties of forming stellar aggregates and clusters.
Using numerical simulations (\S{2}) of decayed and driven supersonic
turbulence in self-gravitating isothermal gas as templates for
molecular cloud dynamics, I discuss the spatial distribution and
timescale (\S{3}) of star formation, the resulting protostellar
accretion rates (\S{4}), and the expected mass spectra for gas clumps
and protostellar cores (\S{5}).

\section{Numerical Method and Driven Turbulence}
\label{sec:numerics}
To adequately describe turbulent fragmentation and the formation of
protostellar cores, it is necessary to resolve the collapse of shock
compressed regions over several orders of magnitude in density. Due to
the stochastic nature of supersonic turbulence, it is not known in
advance where and when local collapse occurs. Hence, SPH ({\em
  smoothed particle hydrodynamics}) is used to solve the equations of
hydrodynamics. It is a Lagrangian method, where the fluid is
represented by an ensemble of particles and flow quantities are
obtained by averaging over an appropriate subset of the SPH particles
(Benz 1990). High density contrasts are resolved by simply increasing
the particle concentration where needed.  SPH can also be combined
with the special-purpose hardware device GRAPE (Sugimoto et al.\ 1990,
Ebisuzaki et al.\ 1993; also Steinmetz 1996) permitting calculations
at supercomputer level on a normal workstation. The simulations
presented here concentrate on subregions within a much larger cloud,
therefore periodic boundary conditions are adopted (Klessen 1997). The
high-density cores of collapsing gas clumps are substituted by `sink'
particles (Bate, Bonnell \& Price 1995) while keeping track of mass
and momentum.
 
The large observed linewidths in molecular clouds imply the presence
of supersonic velocity fields that carry enough energy to
counterbalance gravity on global scales (Williams, Blitz, \& McKee
2000).  However, it is known that turbulent energy dissipates rapidly,
roughly on the free-fall timescale (Mac Low et al.\ 1998, Stone,
Ostriker, \& Gammie 1998, Padoan \& Nordlund 1999). Unlike previously
thought, this is independent of the presence of magnetic fields. For
the current models the fields are therefore assumed to be dynamically
unimportant and not included. To prevent or considerably postpone
global collapse, turbulence is required to be continuously
replenished.  This is achieved by applying a Gaussian driving scheme,
that inserts kinetic energy in a specified range of wavenumbers, $1
\le k \le 2$, $3 \le k \le 4$, and $7 \le k \le 8$, corresponding to
sources that act on large, intermediate, and small scales,
respectively (Klessen et al.\ 2000, models ${\cal B}1h$, ${\cal
  B}2h$, and ${\cal B}3h$).  The energy input at each timestep is
adjusted to reach a constant level of  total kinetic energy,
sufficient to stabilize the cloud as a whole. For comparison, I also
include one molecular cloud model where turbulence already is decayed
and leaves behind a Gaussian density field which begins to contract on
all scales (Klessen \& Burkert 2000, model $\cal I$).

The models presented here are computed in normalized units. If scaled
to mean densities of $n({\rm H}_2) = 10^5\,$cm$^{-3}$, a value typical
for star-forming molecular cloud regions (e.g.\ in $\rho$-Ophiuchus,
see Motte, Andr{\'e}, \& Neri 1998) and a temperature of 11.4$\,$K
(i.e.\ a sound speed $c_{\rm s} = 0.2\,$km/s), then the total mass
contained in the computed volume is 130$\,$M$_{\odot}$ and the size of
the cube is $0.29\,$pc. It contains 64 thermal Jeans masses.

\section{Spatial Distribution and Timescale of Star Formation}
\label{sec:location-time}

Stars form from turbulent fragmentation of molecular cloud material.
Supersonic turbulence that is strong enough to counterbalance
gravity on global scales will usually provoke {\em local} collapse.
Turbulence establishes a complex network of interacting shocks, where
converging shockfronts  generate clumps of high density. This density
enhancement can be large enough for the fluctuations to become
gravitationally unstable and collapse, when the local Jeans
length becomes smaller than the size of the fluctuation.  However, the
fluctuations in turbulent velocity fields are highly transient.  The
random flow that creates local density enhancements can disperse them
again.  For local collapse to actually result in the formation of
stars, locally Jeans-unstable shock-generated density fluctuations
must collapse to sufficiently high densities on time scales shorter
than the typical time interval between two successive shock passages.
Only then are they able to `decouple' from the ambient flow pattern
and survive subsequent shock interactions.  The shorter the time
between shock passages, the less likely these fluctuations are to
survive. Hence, the efficiency of protostellar core formation and the
rate of continuing accretion onto collapsed cores depend strongly on
the wavelength and strength of the driving source
(Klessen et al.\ 2000).

The velocity field of long-wavelength turbulence is dominated by
large-scale shocks which are very efficient in sweeping up molecular
cloud material, thus creating massive coherent structures. When a
coherent region reaches the critical density for gravitational collapse
its mass typically exceeds the local Jeans limit by far.  Inside the
shock compressed region, the velocity dispersion is much smaller than
in the ambient turbulent flow and the situation is similar to
localized tur\-bulent decay. Quickly a cluster of protostellar cores
builds up. Both, decaying and large-scale turbulence therefore lead to a {\em
  clustered} mode of star formation. The efficiency of turbulent
fragmentation is reduced if the driving wavelength decreases. When
energy is carried mainly on small spatial scales, the network of
interacting shocks is very tightly knit, and protostellar cores form
independently of each other at random locations throughout the cloud
and at random times.  Individual shock generated clumps have lower
mass and the time interval between two shock passages through the same
point in space is small.  Hence, collapsing cores are easily destroyed
again and star formation is inefficient. This scenario corresponds to
the {\em isolated} mode of star formation. It needs to be pointed out
that there is no fundamental dichotomy between the two modes of star
formation, they rather define the extreme ends in the continuous
spectrum of the properties of turbulent molecular cloud fragmentation.

\begin{figure*}[t]
\label{fig:a}
\unitlength1cm
\begin{picture}(15.0, 11.7)
\put(0.5,-4.4) {\epsfxsize=12cm \epsfbox{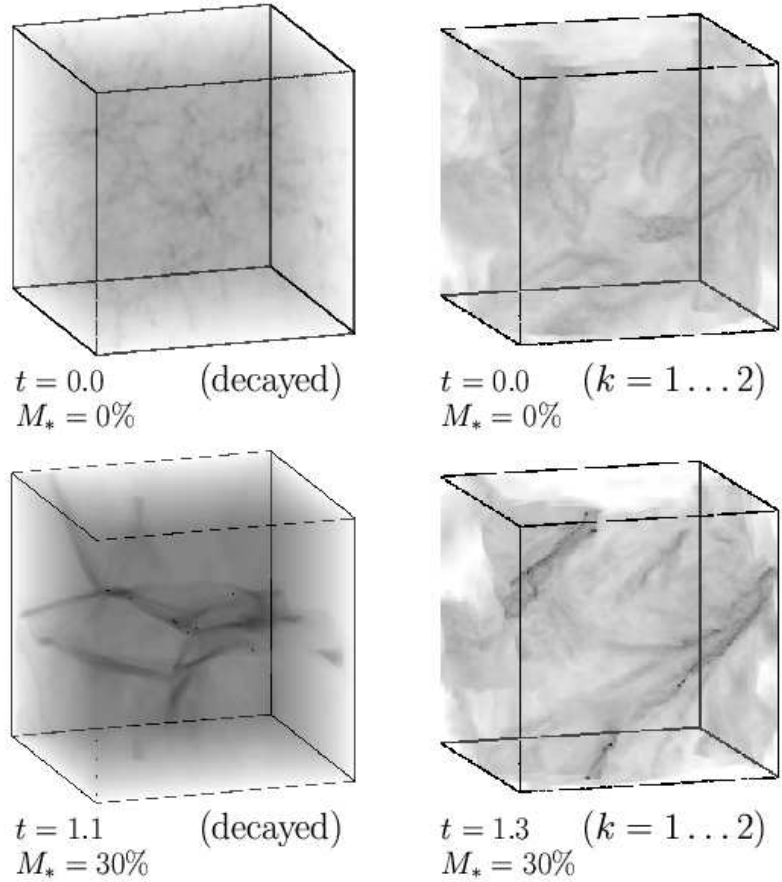}}
\end{picture}
\caption{Comparison of the gas distribution in the models: decayed turbulence
  (left column) and large-scale turbulence (right column).  The upper
  panel depicts the initial stage when gravity is `turned on', the
  lower panel shows system after the first cores have formed and
  accumulated roughly 30\%  of the total mass.
  }
\end{figure*}
\begin{figure*}[t]
\label{fig:a}
\unitlength1cm
\begin{picture}(15.0,11.7)
\put(0.5,-4.4) {\epsfxsize=12cm \epsfbox{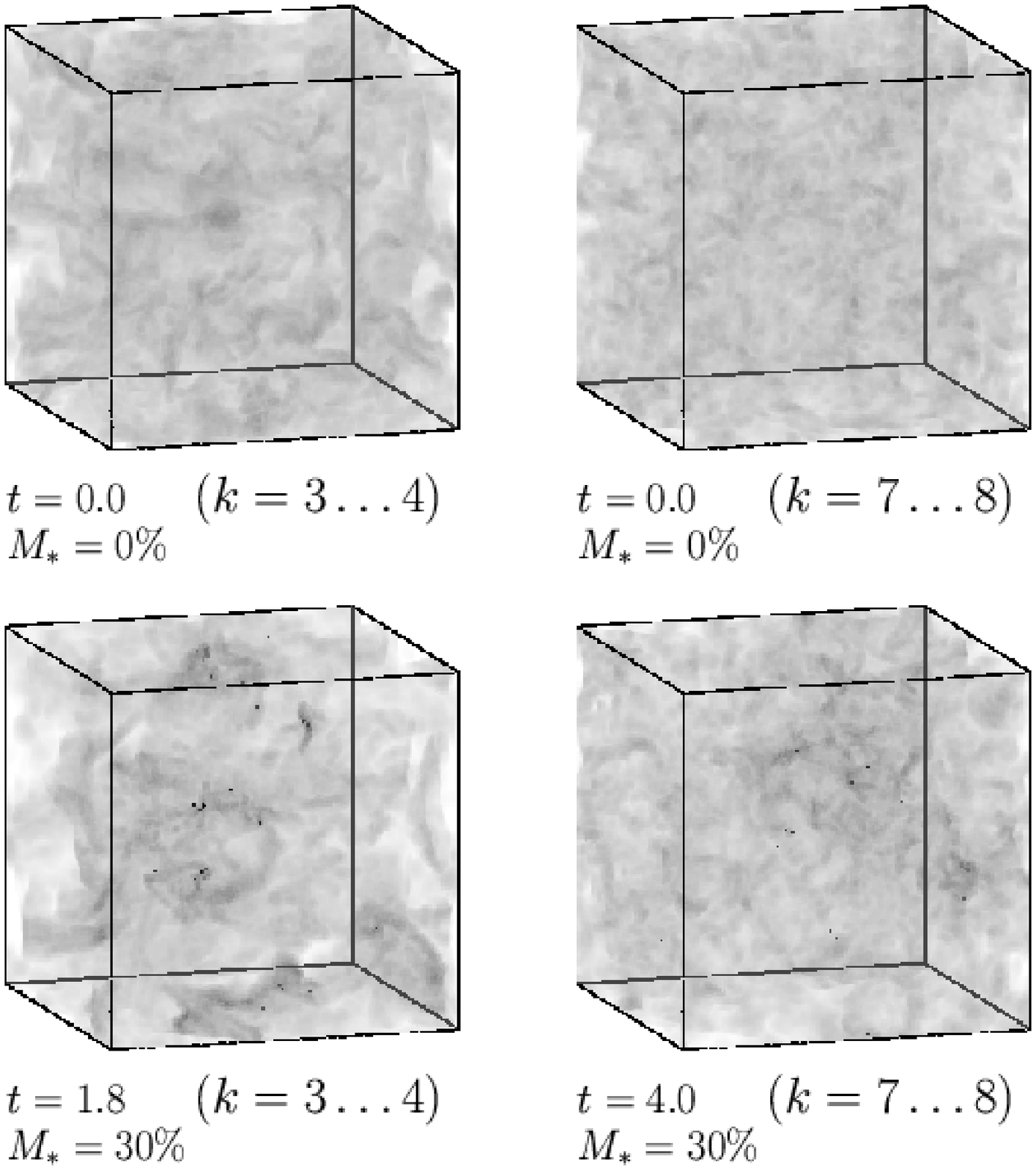}}
\end{picture}
\center{\parbox[t]{11.8cm}{
Figure 1. --- continued: Comparison of the gas distribution in the
models: intermediate-wavelength turbulence (left column) and
small-wavelength turbulence (right column).}}
\end{figure*}

This is visualized in Fig.~1. It compares the models of decayed,
large-wavelength ($k=1\dots2$), intermediate ($k=3\dots4$), and
small-scale turbulence ($k=7\dots8$). The density structure of the
systems is depicted at $t=0$, which means for the model of decayed
turbulence that the initial Gaussian density field is visible. For the
driven models, $t=0$ corresponds to the phase of fully developed
turbulence just before gravity is `switched on'. The lower panel
describes the four models after the first protostellar cores have
formed via turbulent fragmentation and have accreted 30\% of the total
mass.  Time is measured in units of the global free-fall timescale
$\tau_{\rm ff} = (3\pi/32G)^{-1/2}\,\langle\rho\rangle^{-1/2}$, with
$\langle\rho\rangle$ being the mean density. Dark dots indicate the
location of dense collapsed core. As for decayed turbulence all
spatial modes are unstable, the system quickly evolves into a network
of intersecting filaments, where protostellar cores predominantly
form.  Similarly, also large-scale turbulence builds up a network of
filaments, however, this time the large coherent structures are not
caused by gravity, but instead are due to shock compression.  Once
gravity is included, it quickly dominates the evolution inside the
dense regions and again a cluster of protostellar cores builds up. In
the case of intermediate-wavelength turbulence, cores form in small
aggregates, whereas small-scale turbulence leads to local collapse of
individual objects randomly dispersed throughout the volume. Note the
different times needed for 30\% of the mass to be accumulated in dense
cores.  For small-scale turbulence star formation progresses slowest,
but the process speeds up with increasing driving wavelength. The
collapse rates for large-scale turbulence and locally decayed
turbulence are comparable.  This is further exemplified in Fig.~2,
which shows the mass accretion history of individual protostellar
cores for the three turbulent models, together with the distribution
of core formation times. Star formation efficiency is high in the
long-wavelength model, all cores form within two free-fall times,
whereas in the short-wavelength model the efficiency is low and core
formation continues for over 15$\,\tau_{\rm ff}$ (when the simulation
was stopped).
\begin{figure*}[ht]
\label{fig:b}
\unitlength1cm
\begin{picture}(15.0, 9.2)
\put(-1.8,-8.3) {\epsfxsize=16.5cm \epsfbox{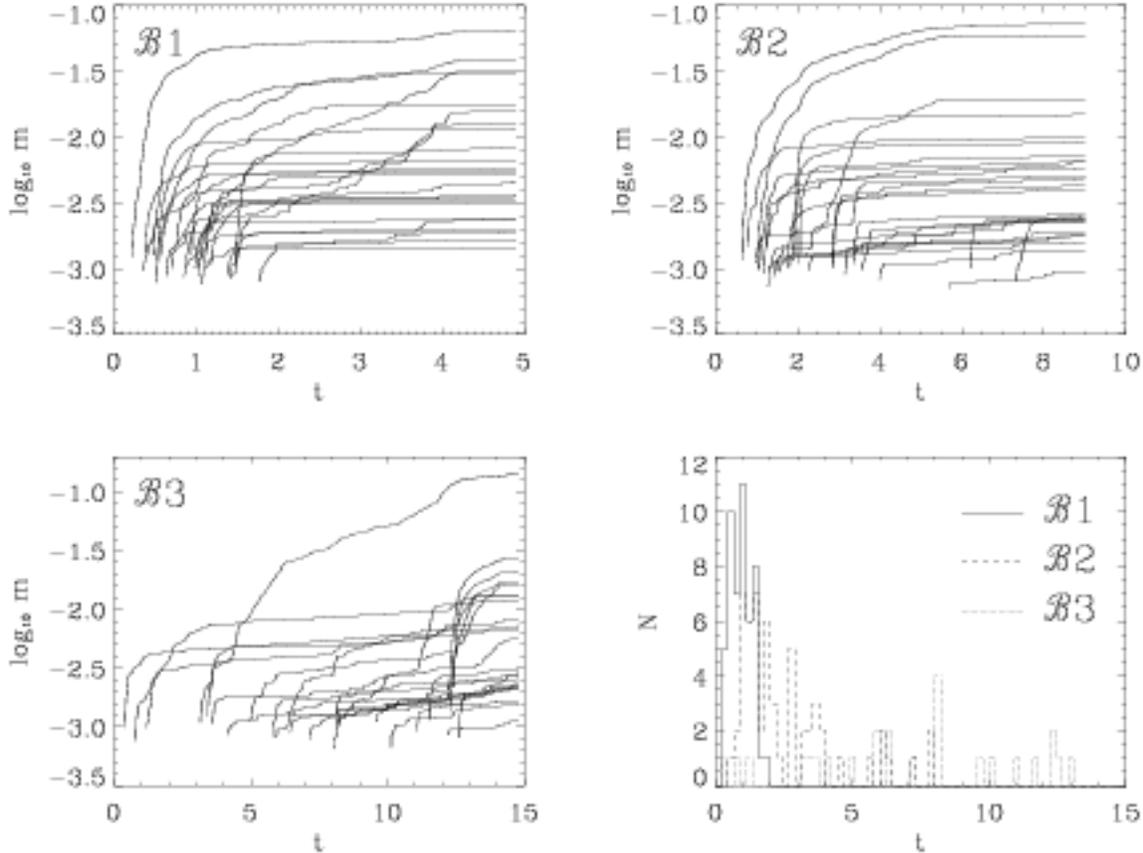}}
\end{picture}
\caption{Mass growth history of protostellar cores (only every
  second core is depicted): long-wavelength (${\cal B}1h$),
  intermediate- (${\cal B}2h$), and small-scale turbulence (${\cal
    B}3h$). The lower left plot shows the distribution of the
  formation times of the cores. Time is again given in units of $\tau_{\rm ff}$
  and masses are scaled to the total mass in the system.  }
\end{figure*}

\section{Protostellar Accretion Rates}
\label{sec:accretion-rates}
\begin{figure*}[p]
\label{fig:b}
\unitlength1cm
\begin{picture}(15.0, 16.5)
\put(-3.4,-1.6) {\epsfxsize=18.5cm \epsfbox{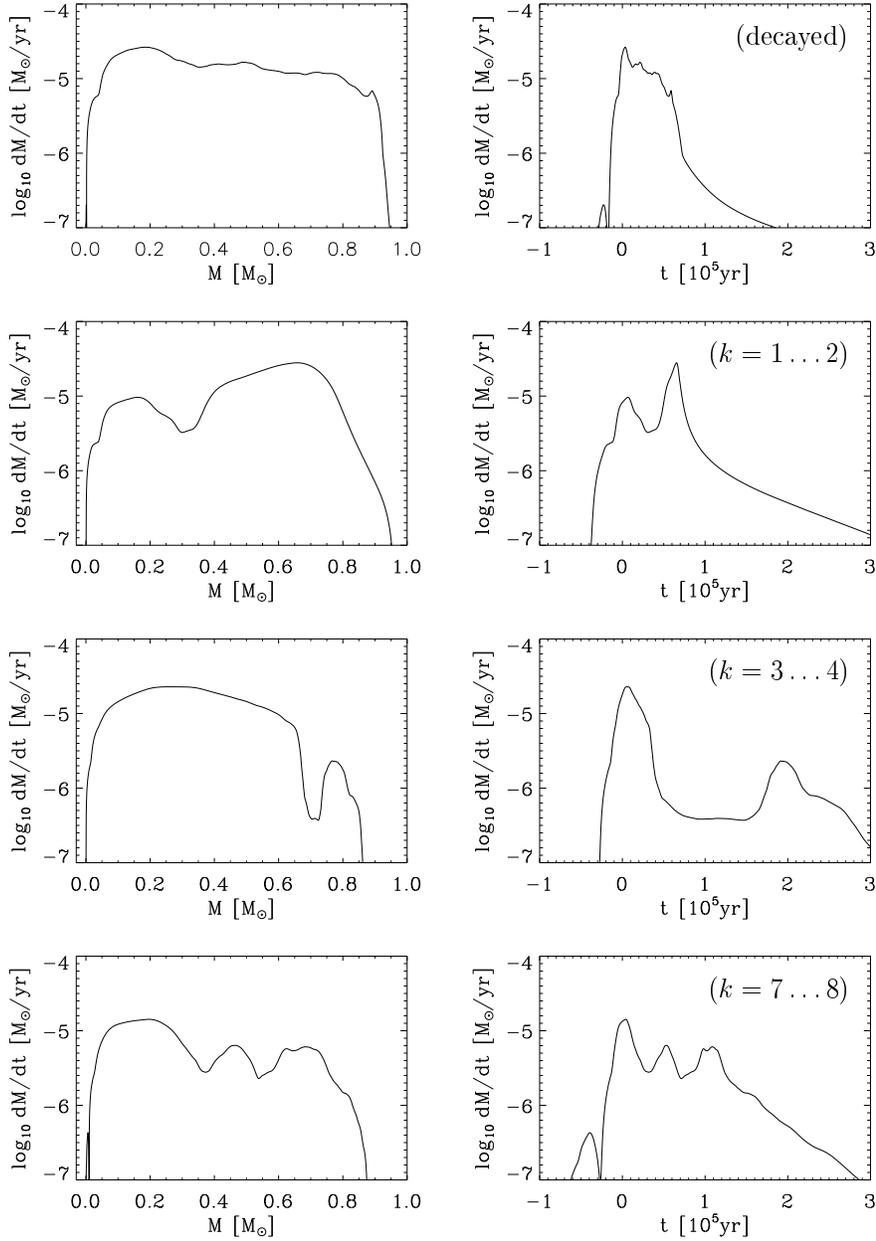}}
\end{picture}
\caption{Accretion rates as function of mass and time for the
  protostellar cores with final mass closest to $1\,$M$_{\odot}$. The
  four models are scaled to a temperature $T=11.4\,$K and a mean density
  $n({\rm H}_2) = 10^5\,$cm$^{-3}$.  Note, that the rate $dM/dt$ is
  independent of the adopted scaling, mass $M$ and time $t$ change
  synchronous. The depicted time interval is chosen such that the
  formation of the protostellar core (`sink' particle) is at $t=0$.
  Values of $dM/dt$ for $t<0$ are due to turbulent compression. }
\end{figure*}

In a dynamical cloud environment the mass accretion rates onto
collapsing protostellar cores are strongly {\em time variable}. This
is already visible in Fig.\ 2, and is quantified in more detail in
Fig.\ 3. It shows $dM/dt$ as function of mass $M$ and time $t$ in each
model for the protostellar core which final mass is nearest to
$1\,$M$_{\odot}$. To obtain mass $M$ and time $t$ in physical units,
the models are scaled to a temperature $T=11.4\,$K and a mean density
$n({\rm H}_2) = 10^5\,$cm$^{-3}$. The accretion rate $dM/dt$ is
insensitive to the adopted scaling, as $M$ and $t$ behave analogously,
e.g.\ if $n({\rm H}_2)$ is decreased by 100, then mass and timescale
go up by a factor 10. The current models yield peak accretion rates of
a few $10^{-5}\,$M$\,$yr$^{-1}$ during the initial phase of
protostellar collapse. For undisturbed cores, the accretion rates then
rapidly decline in the subsequent evolution. This decrease is expected
for dynamical collapse and is suggested by observations (e.g.\ 
Hendriksen, Andr{\'e}, \& Bontemps 1997).  In a dynamical cloud
environment, however, additional peaks of strong accretion are likely
to occur. Their duration and strength are highly probabilistic.  In
the case of large-scale collapse, variations of $dM/dt$ mostly result
from clump mergers which modify the immediate surrounding of the
accreting cores. In turbulent clouds, most secondary accretion peaks
are due to additional shock interaction. Again, gas is compressed in
the vicinity the protostar and the accretion rate increases.

The environmental variation of the accretion rates will impact the
observational properties of forming protostars.  As the accretion
luminosity is proportional to the accretion rate, considerable
deviations are expected between the pre-main sequence evolutionary
tracks of protostars of the same mass and age but different mass
growth histories (see also Wuchterl, this volume).

\section{Mass Spectra of Clumps and Protostellar Cores}
\label{sec:mass-spectra}
The dominant parameter determining stellar evolution is the mass. It
is therefore important to investigate the relation between the masses
of molecular clumps, protostellar cores and the resulting stars.
Figure~4 plots for the four models the mass distribution of all gas
clumps, of the subset of Jeans-critical clumps, and of collapsed
cores. Three different evolutionary phases are shown, initially just
when gravity is `switched on', and then after turbulent fragmentation
has lead to significant protostellar core formation when $M_{\rm
\large *}\approx 30$\% and $M_{\rm \large *}\approx 60$\%,
respectively. In the late stages the gravitational potential is
dominated by the dense cores. Clumps are defined based on a method
proposed by Williams, De~Geus, \& Blitz (1994), which has been 
adapted to the SPH algorithm and makes use of all three spatial
coordinates (see Appendix 1 in Klessen \& Burkert 2000).

In the initial, completely pre-stellar phase the clump mass spectrum
is  very steep (about Salpeter slope or less) at the high-mass end and
gets shallower below $M \approx 0.4 \,\langle M_{\rm J} \rangle$ with
slope $-1.5$, when using a power-law fit to $dN/dM$. The spectrum strongly declines
beyond the SPH resolution limit. Altogether, individual clumps are
rarely more massive than one or two $\langle M_{\rm J}\rangle$.  This
distribution is similar to what Motte et al.\ (1998) find for
pre-stellar condensations in $\rho\,$-Ophiuchus.  Recall that for
densities of $n({\rm H}_2) = 10^5\,$cm$^{-3}$ and temperatures $T =
11.4\,$K, the mean Jeans mass in the system is $\langle M_{\rm J}
\rangle \approx 1\,$M$_{\odot}$. 

Gravity strongly modifies the distribution of clump masses during the
later evolution. As gas clumps merge and grow bigger, their mass spectrum
becomes flatter and extends towards larger masses. Consequently, the
number of clumps that exceed the Jeans limit grows, and local collapse
sets in leading to the formation of dense condensations. This is most
evident in the model of decayed turbulence, where the velocity field
is entirely determined by gravitational contraction on all scales. The
clump mass spectrum in intermediate phases of the evolution (i.e.\ 
when protostellar cores are forming, but the overall gravitational
potential is still dominated by non-accreted gas) exhibits a slope
$-1.5$ similar to the observed one. When the velocity field is
dominated by strong (driven) turbulence, the effect of gravity on the
clump mass spectrum is much weaker. It remains steep, close to or even
below the Salpeter value. This is most clearly seen for
small-wavelength turbulence. Here, the short interval between shock
passages prohibits efficient merging and the build up of a large
number of massive clumps. Only few fluctuations become Jeans unstable
and collapse to form protostars. These form independent of each other
at random locations and times and typically do not interact.
Increasing the driving wavelength leads to more coherent and rapid
star formation. It also results in a larger number of proto\-stars,
which is maximum in the case of pure gravitational contraction
(i.e.\ decayed turbulence).

\begin{figure*}[tp]
\label{fig:c}
\unitlength1cm
\begin{picture}(15.0, 15.6)
\put(-1.6,-0.5) {\epsfxsize=15.4cm \epsfbox{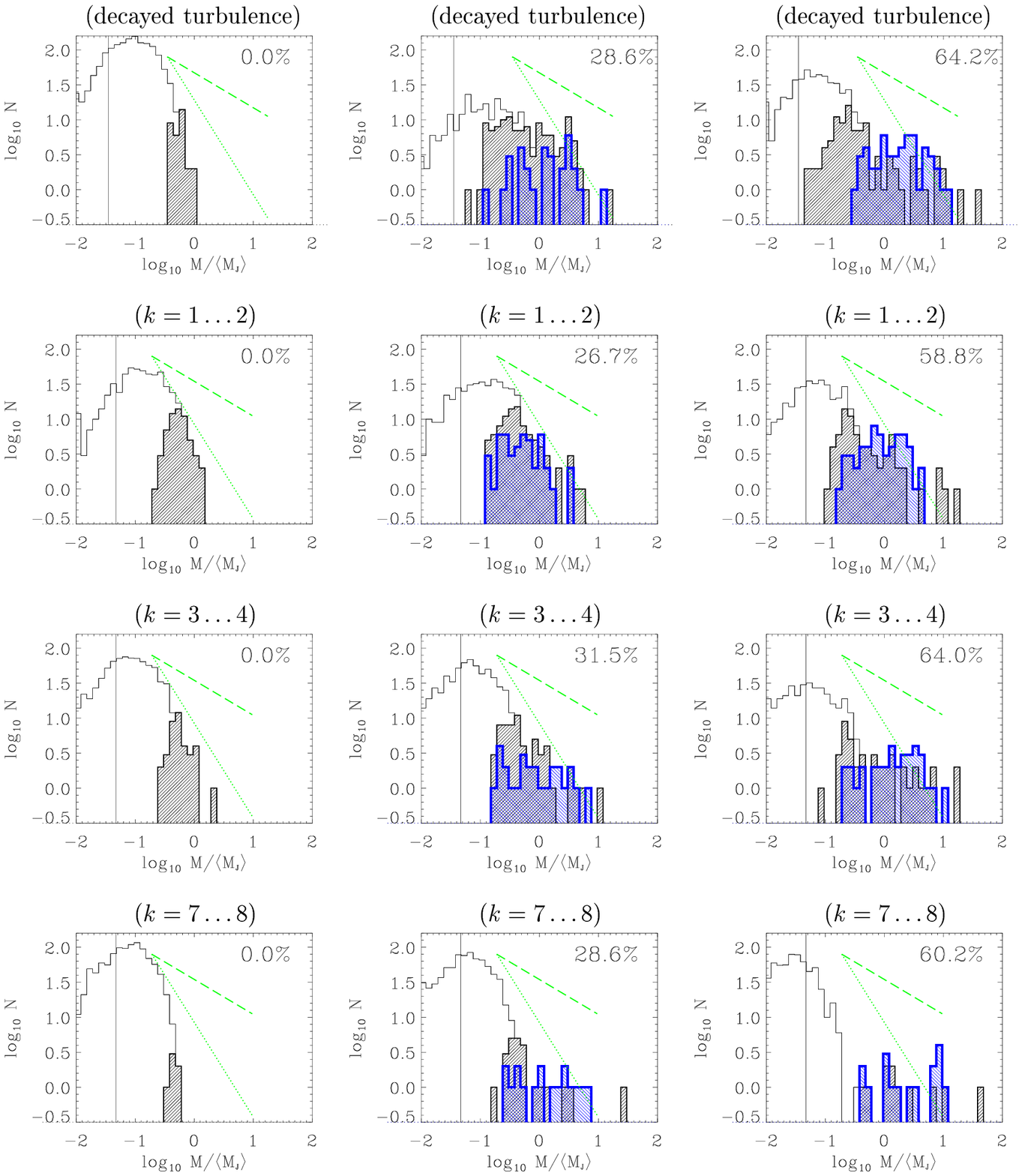}}
\end{picture}
\caption{Mass spectra of dense collapsed cores (hatched
  thick-lined histograms), of gas clumps (thin lines), and of the
  subset of Jeans unstable clumps (thin lines, hatched distribution).
  Masses are binned logarithmically and normalized to the average
  Jeans mass $\langle M_{\rm J}\rangle$. The left column gives the
  initial state of the system, just when gravity is `switched on', the
  middle column shows the mass spectra when roughly 30\% of the mass
  is in dense cores, and the right column when this fraction is about
  60\%. For comparison with power-law spectra ($dN/dM \propto
  M^{\nu}$), the typical slope $\nu = -1.5$ of the observed clump mass
  distribution, and the Salpeter slope $\nu=-2.33$ for the IMF, are
  indicated by the long dashed and by the dotted lines in each plot.
  The vertical line shows the SPH resolution limit.}
\end{figure*}

Long-wavelength turbulence or turbulent decay leads to a core mass
spectrum that is well approximated by a {\em log-normal}. It roughly
peaks at the {\em average thermal Jeans mass} $\langle M_{\rm
  J}\rangle$ of the system and is comparable in width with the
observed IMF (Klessen \& Burkert 2000). The log-normal shape of
the mass distribution may be explained by invoking the central limit
theorem (e.g.\ Zinnecker 1984), as protostellar cores form and evolve
through a sequence of highly stochastic events (resulting from
supersonic turbulence and/or competitive accretion). To find the mass
peak at $\langle M_{\rm J}\rangle$ may be somewhat surprising given
the fact that the local Jeans mass strongly varies between different
clumps. In a statistical sense the system retains knowledge of its
mean properties.  The total width of the core distribution is about
two orders of magnitude in mass and is approximately the same for all
four models.  However, the spectrum for intermediate and
short-wavelength turbulence, i.e.\ for isolated core formation, is too
flat (or equivalently too wide) to be comparable to the observed IMF.
This is in agreement with the hypothesis that most stars form in
aggregates or clusters.

\acknowledgements I thank Andreas
Burkert, Fabian Heitsch, Mordecai-Mark Mac~Low, and G{\"u}nther
Wuchterl for many stimulating discussions and fruitful collaboration.

\end{document}